\documentclass[12pt]{article} 
\usepackage{graphicx}
\usepackage{amsmath,amssymb}

\begin{document}
\baselineskip 21pt
\setcounter{totalnumber}{8}

\bigskip

\centerline{\Large \bf Synchronous Evolution}
\centerline{\Large \bf of Galaxies in Groups: NGC524 Group}

\bigskip

\centerline{\large O.K. Sil'chenko$^1$ and V.L. Afanasiev$^2$}

\noindent
{\it Sternberg Astronomical Institute of the Lomonosov Moscow State University, Moscow, Russia}$^1$

\noindent
{\it Special Astrophysical Observatory of the Russian Academy of Sciences, Nizhnij Arkhyz, Russia}$^2$

\vspace{2mm}
\sloppypar 
\vspace{2mm}

\bigskip

{\small 
\noindent
By means of panoramic spectroscopy at the SAO RAS BTA telescope, we investigated
the properties of stellar populations in the central regions of five early-type 
galaxies -- the NGC 524 group members. The evolution of the central regions of galaxies 
looks synchronized: the average age of stars in the bulges of all the five galaxies 
lies in the range of 3--6 Gyr. Four of the five galaxies revealed synchronized bursts 
of star formation in the nuclei 1--2 Gyr ago. The only galaxy, in which the ages of 
stellar population in the nucleus and in the bulge coincide (i.e. the nuclear burst 
of star formation did not take place) is NGC 502, the farthest from the center of 
the group of all the galaxies studied.
}

\clearpage

\section{INTRODUCTION}

Most galaxies in the nearby Universe are group members: according to recent 
estimates \cite{makkar}, 54\%\ of all the galaxies within 40 Mpc from us are
assembled in groups with four or more members; and if we take into account 
also the pairs and triplets, the number of galaxies in small groups increases 
to 82\%. If considering their properties -- a presence of nearby gravitating 
neighbours and low, about 100 km/s, velocities of mutual passages of galaxies, the
typical dynamic crossing times being several times smaller than the Hubble time --, 
the groups are indeed the best place for the demonstration of effects of so-called 
external secular evolution of galaxies (for the classification of the evolution 
mechanisms of galaxies, see the survey by Kormendy and Kennicutt \cite{kk04}). 
This implies that the external effects, both gravitational and gasdynamical,
must play an important, perhaps even dominant role in the evolution of galaxies 
in groups, by shaping the stellar subsystems of these galaxies. These effects can
dramatically change the structure of galaxies, and lead even to their morphological 
transformation.

The galaxies in groups are affected by the total gravitational potential of the group 
in which they move, and also by the common medium environment (hot intergalactic gas, 
if present in groups). Then, it seems likely that shaping of the global structures 
and, in particular, the formation of central stellar subsystems in group galaxies 
may be synchronized, i.e. the mean ages of the stellar populations in the centers of 
galaxies would be similar. Almost all the mechanisms of the gravitational (tidal) 
interaction, as well as many gasdynamical mechanisms (ram pressure, static compression 
by the hot medium) lead to the gas concentration in the galactic centers, which provokes
subsequent starbursts resulting in formation of secondary (now intermediate-age) 
stellar population in the nucleus and over the bulge of the galaxy. The discovery 
of the synchronization in the evolution of galactic central regions could
be an argument in favour of the dominant role of external effects, and the lack 
of the sinchronization, on the contrary, should attract attention to the differences of
initial conditions in the evolution of galaxies.

More than ten years ago we started study of stellar populations in central 
regions of galaxies in groups with the Multi-Pupil Fiber Spectrograph (MPFS) of the
Russian 6-m telescope of the Special Astrophysical Observatory of the Russian
Academy of Sciences (SAO RAS). Up to now we have investigated several, typically 
2--3 central galaxies in six nearby groups. In Leo I \cite{leo1}, and in the 
NGC 5576 \cite{n5574} and NGC 3169 \cite{n3169} groups of galaxies the properties of
stellar population in the circumnuclear disks of large galaxies turned out to be 
the same. The last star formation there has stopped relatively recently, 1–3 Gyr ago,
in spite of the early type of the host galaxies, and in the Leo I group 
the circumnuclear stellar disks even have the same spatial orientation of the
rotation axes. In the Leo triplet (NGC3623/NGC 3627/NGC 3628), on the
contrary, the age of the stellar population and the kinematics of the gas 
in the centers of NGC 3623 and NGC 3627 are significantly different \cite{leo3}. 
We hence concluded that the galaxies of the triplet have assembled into a group 
only recently, less than 1 Gyr ago. In the Leo II group, where unlike in
the previous cases, the X-ray emitting gas was detected around two central galaxies, 
the observed circumnuclear kinematically decoupled stellar substructures turned 
out to be old. They have different ages -- 6 Gyr in NGC 3608 and 10 Gyr in NGC 3607 
\cite{leo2}. At the same time, the peripheral lenticular galaxies NGC 3599 and
NGC 3626 reveal the very young age and even locally current star formation in
the central stellar subsystems \cite{leo2_s0}. The evolution of the central
regions of early-type galaxies in the rich massive group NGC 80 also turned out 
to be asynchronous. This group has a large X-ray halo and is sometimes classified
as a poor cluster. In this group the early-type galaxies have very different SF
histories in their central parts: estimates of the mean age of stellar populations 
both in the nuclei and in the central spheroids are ranging from 1.5--3 Gyr (in the
giant elliptical galaxy NGC 83 and in the S0 galaxy IC 1548) to more than 10 Gyr 
in NGC 79 (an elliptical) and IC 1541 (a lenticular galaxy).

This paper presents the results of analysis of the properties of stellar population 
in the galaxies of another massive group with X-ray emitting gas: the NGC 524 group. 
This group was cataloged for the first time by Geller and Huchra in 1983 \cite{gellerhuchra}. 
Then, only 8 member galaxies were listed in it. Later, Jaan Vennik \cite{vennik} has 
made a visual inspection over the Palomar maps and found 18 bright and 13 dwarf galaxies 
there via the interaction hierarchy method. As concerning the latest catalogs, 
Brough et al. \cite{brough}, by applying the FOF (friend-of-my-friend) method, 
have identified 16 members in the NGC 524 group, and, according to the Catalog
of Nearby Groups by Makarov and Karachentsev \cite{makkar}, the group also contains 
16 member galaxies. If we adopt the radius of the group from the catalog
by Makarov and Karachentsev, $R_h = 391$~kpc, the NED retrieves in this group
10 galaxies brighter than $M_B \sim ?16$. Most of them are classified as lenticular
galaxies. Given the galaxy velocity dispersion in the group of about 150–190 km/s \cite{makkar,brough}, 
the group must be rather massive, $1.7 \times 10^{13} M_{\odot}$ \cite{gems}. The ROSAT
mapped the X-ray emitting gas in the NGC 524 group, however, the radius of the X-ray
spot is less than 60 kpc, hence in some papers the hot gas is not considered as belonging 
to the halo of the group, but only to the central galaxy NGC 524 \cite{xray}.
By undertaking the panoramic spectroscopy we have investigated stellar populations 
and kinematics of circumnuclear regions in the brightest non-central galaxies of 
the group: NGC 502, NGC 509, NGC 516 (all lenticulars), NGC 518, and NGC532 (both 
early-type spirals). The global characteristics of the galaxies under consideration, 
adopted from the NED and HYPERLEDA databases, are given in Table 1. As for the central 
galaxy of the group, NGC 524, the results for it obtained with the MPFS/BTA have been
published by us earlier \cite{n524}.

{\small
\begin{table*}
\caption[ ] {Global Parameters of the Galaxies}
\begin{flushleft}
\begin{tabular}{lccccc}
\hline\noalign{\smallskip}
Galaxy & NGC~502 & NGC~509 & NGC~516 & NGC~518 
&  NGC~532 \\
Type (NED$^1$) & SA$0^0$(r) & S0? & S0 & Sa: &
 Sab? \\
$R_{25},\, ^{\prime \prime}$  (RC3$^2$) & 34 & 43 & 39 & 52 
& 75 \\
$B_T ^0$ (RC3) & 13.57 & 14.20 & 13.97 & 13.56  & 12.91 \\
$M_B$ (RC3$+$NED) & --18.3 & --17.7 & --17.9 & --18.3 & --19.0 \\
$V_r$, km/s (NED) & 2489 & 2274 & 2451 & 2725 & 2361 \\
Separation off the group center$^{3,2}$, kpc & 282 & 151 & 68 & 101 & 126 \\
Distance to the group$^4$, Mpc & \multicolumn{5}{c}{24} \\
\hline
\multicolumn{6}{l}{Photometric parameters of the bulges, according to \cite{n524phot}}\\
$r_{eff},\, ^{\prime \prime}$ & 3.5 & 2.9 & 2.9 & 7.4 & 5.4 \\
$n_{Sersic}$ & 1.5 & 1.5 & 1.85 & 2 & 2 \\
\multicolumn{6}{l}{Kinematical parameters of the bulges, according to the present work} \\
$V_{rot}(r_{eff})$, km/s & 59 & $\sim 20$ & $\sim 30 $ & 57 & $\sim 40$ \\
$\sigma _* (r_{eff})$, km/s & 96 & 77 & 65 & 98 & 110 \\
\hline
\multicolumn{6}{l}{$^1$\rule{0pt}{11pt}\footnotesize
NASA/IPAC Extragalactic Database}\\
\multicolumn{6}{l}{$^2$\rule{0pt}{11pt}\footnotesize
Third Reference Catalogue of Bright Galaxies \cite{rc3}}\\
\multicolumn{6}{l}{$^3$\rule{0pt}{11pt}\footnotesize
from \cite{sengupta}}\\
\multicolumn{6}{l}{$^4$\rule{0pt}{11pt}\footnotesize
from \cite{tonry}}\\
\end{tabular}
\end{flushleft}
\end{table*}
}

\section{OBSERVATIONS AND DATA REDUCTION}

Panoramic spectral data were obtained for the central parts of five large 
galaxies of the NGC 524 group with the Multi-Pupil Fiber Spectrograph (MPFS) 
installed in the primary focus of the 6-m BTA telescope of the SAO RAS --
for the description of the instrument, see \cite{mpfs}. To study the stellar 
populations and stellar kinematics, the blue--green spectral range of 4150-5650~\AA\ 
was observed with reciprocal dispersion of 0.75~\AA\ per pixel (spectral resolution
of about 3~\AA). For the galaxy with an emission-line spectrum (with a large amount 
of warm ionized gas in the center), NGC 532, we also exposed the red spectral range 
5800–7200~\AA\ to investigate the kinematics of the gas component in the
center of the galaxy, and also to calculate the correction to the Lick index 
H$\beta$ for the contamination of the absorption line by emission by using the 
measurement of the equivalent width of the H$\alpha$ emission line. The detector 
used was a EEV42-40 CCD, with a format of $2048 \times 2048$ pixels. At the 
observations with the MPFS, an array of square microlenses, assembled
into a square field of view sized $16 \times 16$ elements constructs a matrix 
of micro-pupils, which is then reformed into a pseudo-slit via the optical fibers 
and fed to the spectrograph input. This configuration allows to record 256 spectra
in one exposure, each of which characterizes a spatial element of the image
of the galaxy sized about $1'' \times 1''$. To calibrate the wavelength scale, 
each time before and after the target, a comparison spectrum of the helium--neon--argon
lamp was exposed; in order to correct for the vignetting and different transmissions 
of the microlenses, we used the inbuilt continuum lamp of the spectrograph and the
spectrum of the twilight sky. The primary data processing, which includes the bias 
subtraction, removal of traces of cosmic hits, extraction of one-dimensional spectra 
for every pixel, and linearization of the extracted spectra, was performed using 
original software written in the IDL environment.

To calculate the kinematical parameters of the stellar component -- line-of-sight velocities 
and stellar velocity dispersions -- the spectrum of each spatial element after the continuum 
subtraction and the transformation to the velocity scale was cross-correlated
with the spectra of type G8--K2 giant stars. They were observed the same night and using
the same instruments as the galaxies. For the cross-correlation of spectra of younger stellar 
populations we used the twilight sky spectrum (of G2-type) as the reference frame. The
accuracy of the wavelength scales and the zero point of measured velocities were checked with 
the night-sky line [OI]$\lambda$5577~\AA. We estimate the accuracy of individual measurements 
of velocities for the stellar component as 5--7 km/s. After having obtained the kinematical 
parameters, we derived the properties of the stellar populations -- their age, total metallicity
and the ratio of the abundances of magnesium (the $\alpha$--process product) and iron -- through 
the Lick indices H$\beta$, Mgb, Fe5270, and Fe5335 for every one-element spectrum. 
Definition of the Lick index system was taken following the papers by Worthey et al. 
\cite{woretal94,worjones}. The resulting maps of the Lick indices -- Mgb and 
$\langle \mbox{Fe} \rangle \equiv (\mbox{Fe5270} + \mbox{Fe5335})/2$ for all galaxies
and H$\beta$ for the lenticular galaxies without noticeable emission lines in the spectra --
are presented in Figs. 1--5. Detailed model calculations exist for these strong absorption 
lines within the stellar population synthesis models, which associate the indices 
(equivalent widths of absorption lines) with the mean (weighted with the luminosity
of stars) properties of the stellar populations, the so--called SSP--equivalent 
(SSP$=$Simple or Single Stellar Population) ages and metallicities. From the variety 
of available model calculations, we have chosen the Thomas et al. models \cite{thomod}
for the comparison with our observational results, since these models are calculated 
for several values of the magnesium-to-iron abundance ratio, and, therefore,
provide determination, besides the age and total metallicity, of this third parameter 
of the stellar population which is very important to restrict the duration of the star
formation. A detailed log of observations of the NGC 524 group galaxies 
is presented in Table 2.

\begin{table*}
\caption[ ] {Log of the spectral observations with the MPFS}
\begin{flushleft}
\begin{tabular}{llccc}
\hline\noalign{\smallskip}
Galaxy & Date &  Exposure, min & Spectral range, \AA & Seeing, arcsec \\
\hline\noalign{\smallskip}
NGC 502 & 17.09.2006 & 60  & 4150--5650 & 2.0 \\
NGC 509 & 18.09.2006 & 120 & 4150--5650 & 1.5 \\
NGC 516 & 17.09.2006 & 60 & 4150--5650 & 2.0 \\
NGC 518 & 19.08.2007 & 150 & 4150--5650 & 1.3 \\
NGC 532 & 08.10.2004 & 120 & 4150--5650 & 1.2 \\
NGC 532 & 06.09.2008 & 40 & 5800--7300 & 1.5 \\
\hline
\end{tabular}
\end{flushleft}
\end{table*}

\section{RESULTS}

\subsection{Two-dimensional distributions of Lick indices}

The morphology of the stellar population parameters in the central 
regions of galaxies can be qualitatively characterized by the distributions
of the Lick indices of metals (magnesium and iron) and hydrogen.
These distributions are shown in Figs. 1–5.
Let us consider individual features of these distributions for every galaxy.

\begin{figure*}[!h]
\includegraphics[scale=0.8]{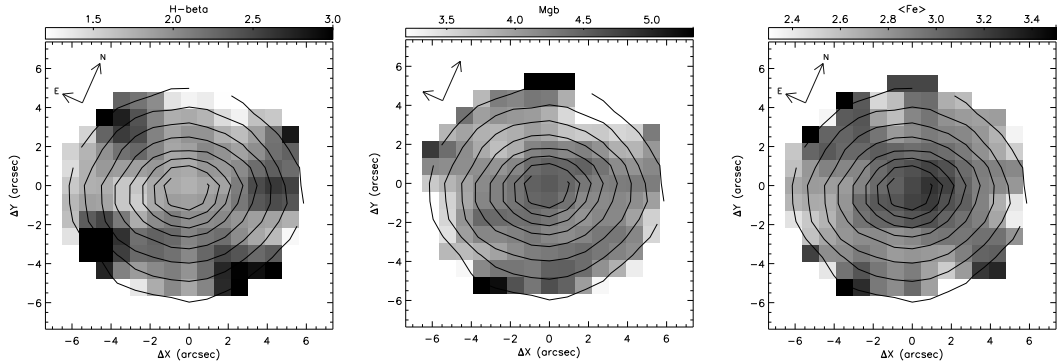}
\caption{
The maps of the Lick indices H$\beta$, Mgb, and $<\mbox{Fe}> \equiv$(Fe5270+Fe5335)/2
for the central region of NGC~502 according to the MPFS data. The isophotes
show the distribution of surface brightness in the green ($\lambda$5000~\AA) continuum.}
\label{n502}
\end{figure*}

\noindent
{\bf NGC~502, Fig.~1.}  The morphology of the Lick indices
in NGC~502 can be described as an increase of the equivalent line widths 
of magnesium and iron in the nucleus with a simultaneous weakening of H$\beta$.
This can result from an enhanced metallicity of stars in the geometric center 
of the galaxy: the galaxy reveals a chemically decoupled nucleus. In
other respects, no noticeable features or anisotropy are observed in the distributions: 
NGC~502 is viewed almost face-on, and the properties of the stellar population
are centrally symmetric.

\begin{figure*}[!h]
\includegraphics[scale=0.8]{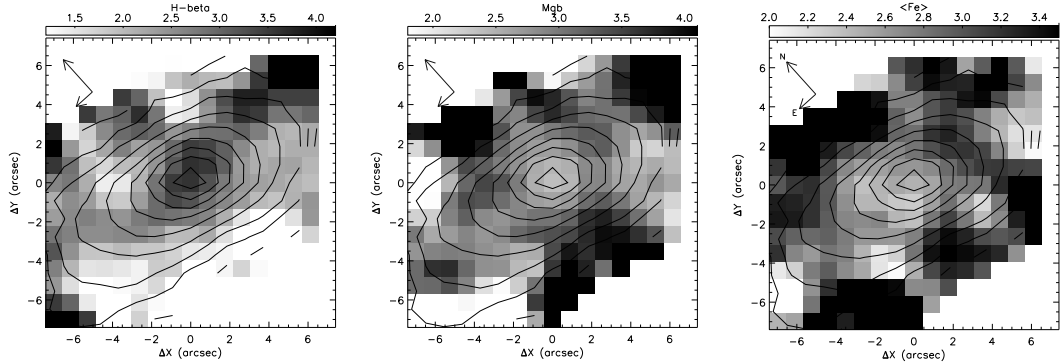}
\caption{
The maps of the Lick indices H$\beta$, Mgb, and $<\mbox{Fe}> \equiv$(Fe5270+Fe5335)/2
for the central region of NGC~509 according to the MPFS data. The isophotes
show the distribution of surface brightness in the green ($\lambda$5000~\AA) continuum.}
\label{n509}
\end{figure*}

\noindent
{\bf NGC~509, Fig.~2.} This lenticular galaxy, viewed edge-on, demonstrates an opposite 
character of the distributions of Lick indices: one can note a prominent peak of
H$\beta$ in the nucleus, a less pronounced enhancement of the hydrogen index near 
the equatorial plane of the galaxy, as well as the compact minima of the magnesium
and iron indices near the center of the isophotes. We will see below that such morphology 
mainly comes from the effect of age: the stellar population in the nucleus is much
younger than that around the nucleus. Interestingly, the Mgb minimum is embedded 
into a nearly circular ring of high Mgb index values, although the isophotes
are elongated much stronger than the `isoindex' contours. It justifies the averaging
of the indices in round rings to estimate the stellar population parameters in the bulge.

\begin{figure*}[!h]
\includegraphics[scale=0.8]{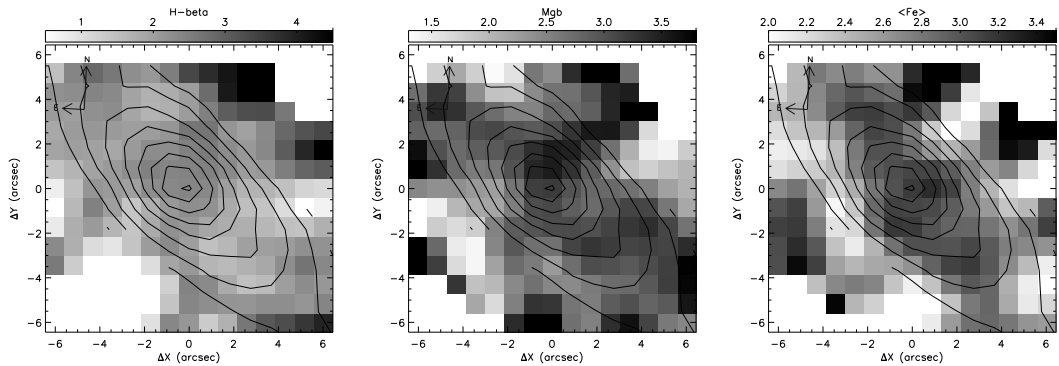}
\caption{
The maps of the Lick indices H$\beta$, Mgb, and $<\mbox{Fe}> \equiv$(Fe5270+Fe5335)/2
for the central region of NGC~516 according to the MPFS data. The isophotes
show the distribution of surface brightness in the green ($\lambda$5000~\AA) continuum.}
\label{n516}
\end{figure*}

\noindent
{\bf NGC~516, Fig.~3.} NGC~516 is a small lenticular galaxy, like NGC~509, 
and as well is viewed edge-on. It is located very close to the center of the group, 
only about 70 kpc off the central group galaxy NGC~524 in projection onto the sky plane, 
and their systemic velocities are also close. In Fig.~3, just like as in the case of NGC~509,
we observe the strengthening of the H$\beta$ index in the nucleus, but this time the indices 
of magnesium and iron exhibit the maxima instead of minima at this point. Moreover, 
the increase of the iron index clearly has a disk morphology: it is not a compact
spot, but rather the narrow lane across the entire field of view. If the increase of H$\beta$ 
is an effect of the age (as we shall see below), we can conclude that the starforming burst 
in NGC~516 was more prolonged than that in the center of NGC~509, that the star formation
was not confined to the nucleus, but rather was expanded over the disk, and due to it 
the stellar disk managed to get enriched with heavy elements.

\begin{figure*}[!h]
\includegraphics[width=0.8\hsize]{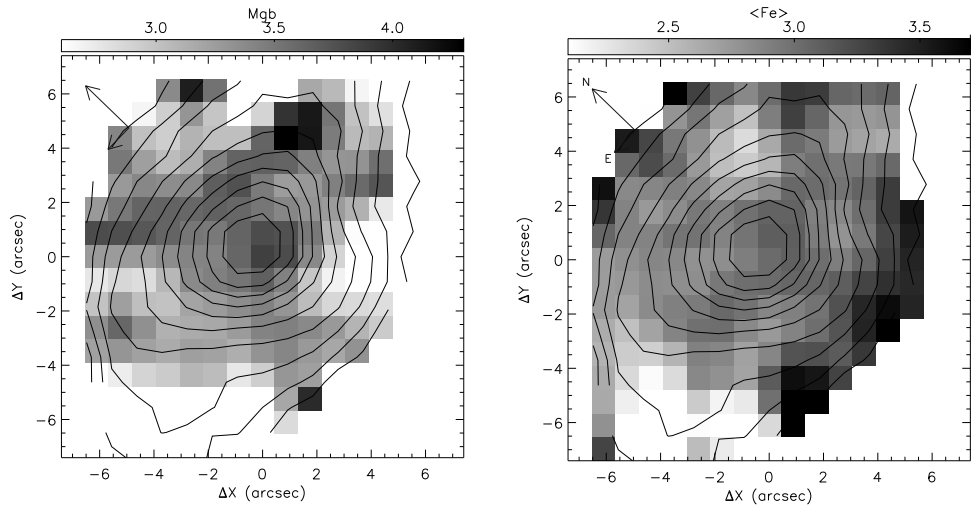}
\caption{The maps of the Lick indices Mgb and $<\mbox{Fe}> \equiv$(Fe5270+Fe5335)/2
for the central region of NGC~518 according to the MPFS data. The isophotes
show the distribution of surface brightness in the green ($\lambda$5000~\AA) continuum.}
\label{n518}
\end{figure*}

\noindent
{\bf NGC~518, Fig.~4.}The spiral galaxy NGC~518 is viewed under a high inclination, 
but not exactly edge-on \cite{n524phot}. The spectra of the central region
reveal a moderate emission line at the bottom of the H$\beta$ absorption line, 
hence in this case the Lick index H$\beta$ is contaminated by emission, and 
in Fig.~4 we do not give its map because it is not informative concerning the age.
Before using it to analyze the properties of stellar population, it has to be 
corrected for the contribution of the emission. A chemically decoupled stellar 
nucleus of the galaxy can be distinguished in the distributions of the magnesium 
and iron indices; it is possibly compact only in Mgb, and has an extended morphology
in the iron index distribution. We encountered this feature at the centers of early-type 
galaxies more than once -- e.g., in the S0-galaxy NGC~1023 \cite{n1023}. It can be 
interpreted as a rapid exhaustion of the star formation burst in the nucleus and 
a subsequent continuous process of star formation in the circumnuclear ring.

\begin{figure*}[!h]
\includegraphics[width=0.8\hsize]{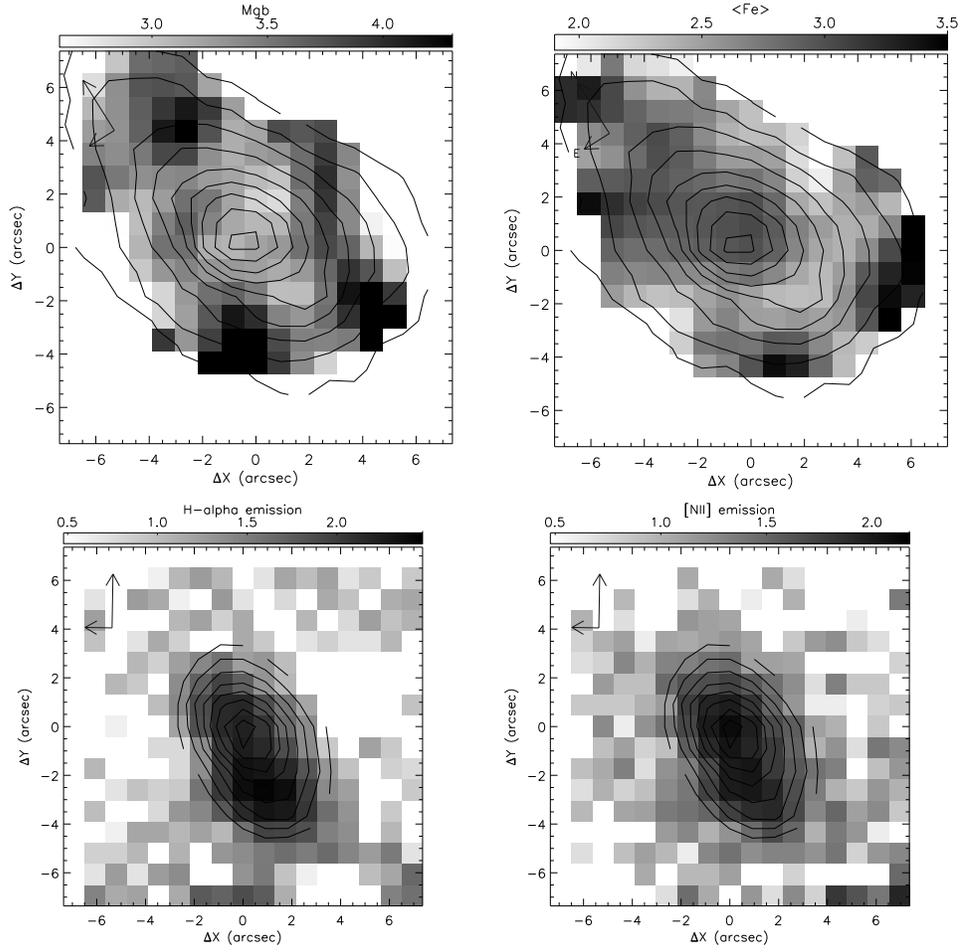}
\caption{The maps of the Lick indices Mgb and $<\mbox{Fe}> \equiv$(Fe5270+Fe5335)/2 ({\it top})
and of the fluxes in the emission lines H$\alpha$ and [NII]$\lambda$6583, in arbitrary units, 
({\it bottom}) for the central region of NGC~532 according to the MPFS data. The isophotes
show the distribution of surface brightness in the green ($\lambda$5000~\AA\, top) and
red ($\lambda$6500~\AA\, bottom) continuum.}
\label{n532}
\end{figure*}

\noindent
{\bf NGC~532, Fig.~5.} In the central region of this galaxy the emission lines of warm ionized 
gas are very strong. To properly take them into account, we took an MPFS spectrum in the 
red range, and in Fig.~5, besides the maps of the Lick indices of magnesium and iron 
(which are equivalent widths of absorption lines), we additionally present the maps of 
flux distribution in the H$\alpha$ and [NII]$\lambda$6583 emission lines (in arbitrary 
counts). All the distributions look complex and asymmetrical, even the maps of the 
emission-line flux distributions. The peak of emission of the forbidden nitrogen line 
comes from the nucleus, whereas the maximum of the H$\alpha$ 
emission line -- from a compact region of about $4^{\prime \prime}$ to south of the nucleus. 
From the flux ratio in these lines (H$\alpha$/[NII] = 1.66 in the nucleus, and above 2 outside 
the nucleus), using the diagnostics by Veilleux and Osterbrock \cite{vo87}, we can say
that the excitation of the nucleus of NGC~532 is a LINER-like, and the major part of current 
star formation is localized in the circumnuclear disk, and exactly to the south of the
nucleus. The measured emission line intensities and the character of excitation of the 
ionized gas we have determined, is used later on to correct the Lick H$\beta$ index for 
the contamination by emission.

\subsection{Age and chemical composition of stellar populations in the nuclei and bulges}

To employ benefits of the panoramic spectroscopy, we summarized the spectra in the rings, 
centered onto the nuclei of the galaxies, when advancing along the radius step by step of
one arcsecond. This has allowed to obtain the spectra for the galaxy bulges with the 
same signal-to-noise ratio, as for the nuclei. With our exposures of about one hour 
and after the summation of the spectra in the rings, the accuracy of measured indices 
is about 0.1~\AA, that was determined by us during the larger survey of lenticular
galaxies with the MPFS through the repeated observations of some galaxies \cite{lenssum}. 
Further we discuss the characteristics of the stellar populations in the nuclei and 
in the ring zones of the bulges, which we have obtained from the Lick indices.

To determine the average (weighted with the luminosity of stars) characteristics 
of the stellar populations: the age, total metallicity, the abundance ratio 
of magnesium and iron, let us compare our measurements of the Lick indices in the
galaxies with the models from \cite{thomod} calculated in the framework of the SSP 
(Simple or Single Stellar Population) star formation histories, that is, assuming 
the same age and the same chemical composition in all stars that contribute to
the integrated spectrum (of a single spatial element of the galaxy in our case). 
To determine the abundance ratio of the magnesium and iron, we will use an index-index
diagram that confronts the Mgb and $<\mbox{Fe}> \equiv$(Fe5270+Fe5335)/2 indices. To
determine the age and total metallicity we compare the H$\beta$ hydrogen index and 
a composite index of metals, [MgFe]$\equiv \sqrt{\mbox{Mgb}\langle \mbox{Fe} \rangle}$: 
this diagram allows to separate the effects of age and metallicity (to solve
the degeneracy), since the sequences of equal metallicity and equal age have 
different slopes and form a kind of a curvilinear coordinate grid, which
allows to determine both the age and metallicity of the stellar population based 
on the position of the observational point within this grid. The H$\beta$ index, which
is used to determine the age of the stellar population should be corrected for the contamination 
by the emission component. In the case of NGC~532, for which we have got a spectrum in the red 
range, this was done through the equivalent width of the H$\alpha$ emission line. 
In the nucleus, we used the average ratio of the equivalent widths of the Balmer emission 
lines EW(H$\beta _{emis}) = 0.25$EW(H$\alpha _{emis})$, found from a large inhomogeneous
sample of spiral galaxies by Stasinska and Sodre \cite{ss01}. At a distance of 4''-6'' 
from the nucleus, where current star formation dominates as a gas excitation mechanism, 
instead of 0.25 we used a coefficient of 0.35, typical for the gas excitation by young 
stars \cite{osterbrock}. If we use the relation from Stasinska and Sodre \cite{ss01} for the
bulge, the age estimate for the bulge would decrease by 0.5 Gyr -- from 4 to 3.5 Gyr. 
For NGC~518, where the emission lines are weaker than in NGC~532, and there exist no data 
in the red spectral range, we used a statistical mean ratio 
EW(H$\beta ) = 0.6$EW([OIII]$\lambda$5007~\AA) 
found  for the early-type galaxies by \cite{trager00a}. The calculated in such a way 
equivalent width of the H$\beta$ emission line is a correction to be added to the
measured H$\beta$ index, to correct it for the emission (by the definition of the 
Lick indices as the equivalent widths).

\begin{table*}
\caption[ ] {Lick indices and stellar population parameters in the nuclei}
\begin{center}
\begin{tabular}{l|c|c|c|c|c|c|}
\hline\noalign{\smallskip}
Galaxy & H$\beta$ &  Mgb & $\langle \mbox{Fe} \rangle $ 
& T, Gyr & [Z/H] & [Mg/Fe] \\
\hline\noalign{\smallskip}
NGC 502 & 1.83 & 4.60 & 3.29 & 3.5 & $+0.6$ & $+0.13$  \\
NGC 509 & 3.99 & 1.85 & 2.00 & 1 & 0.0 & 0.0 \\
NGC 516 & 2.49 & 3.08 & 3.00 & 2 & $+0.4$ & 0.0 \\
NGC 518 & 2.36 & 3.86 & 2.84 & 2 & $+0.5$ & $+0.17$ \\
NGC 532 & 3.07 & 3.19 & 2.99 & 1 & $>+0.7$ & 0.0 \\
\hline
\end{tabular}
\end{center}
\end{table*}

\begin{figure*}[h]
\begin{center}
\begin{tabular}{c c}
 \includegraphics[width=6cm]{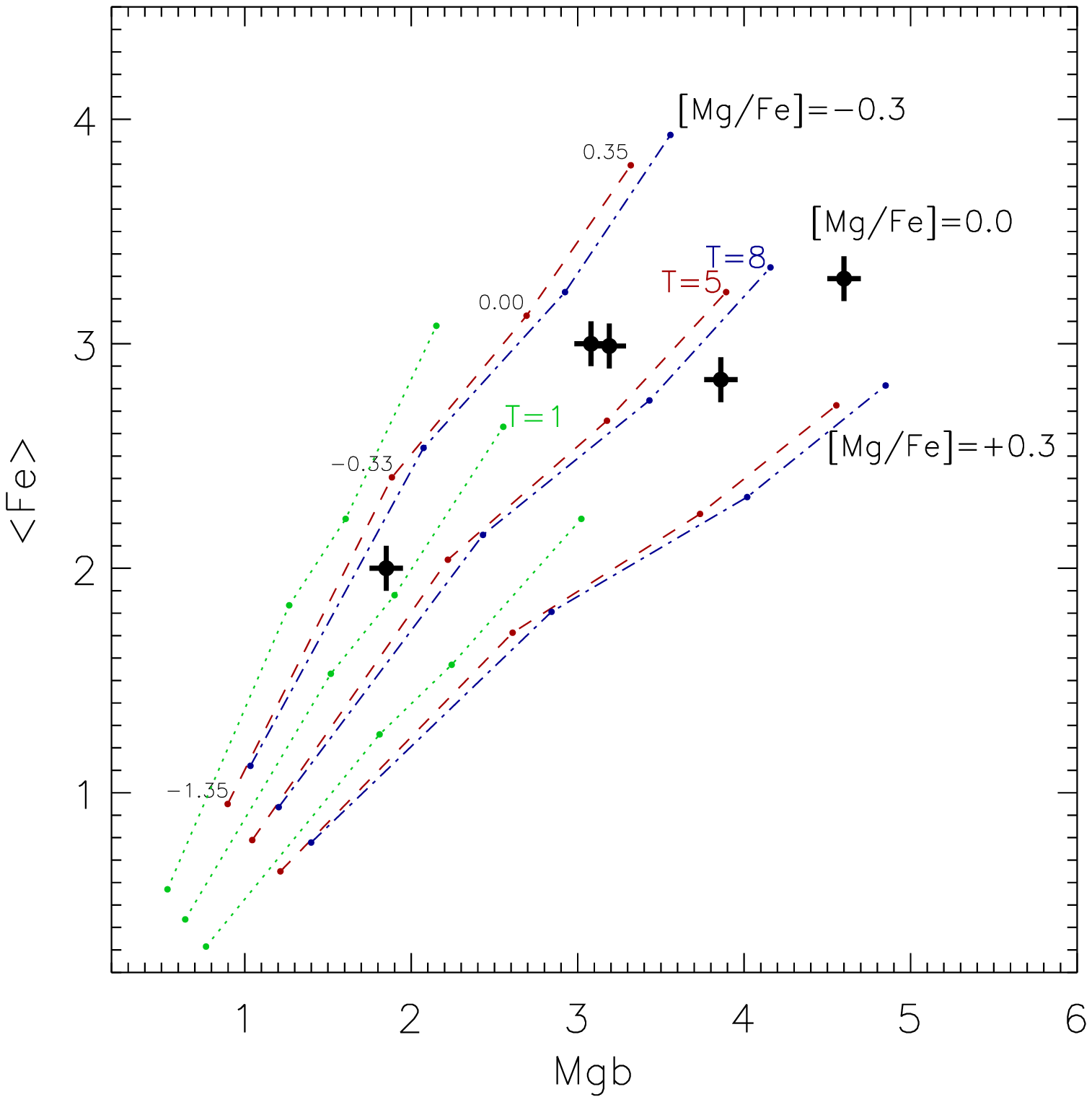} &
 \includegraphics[width=6cm]{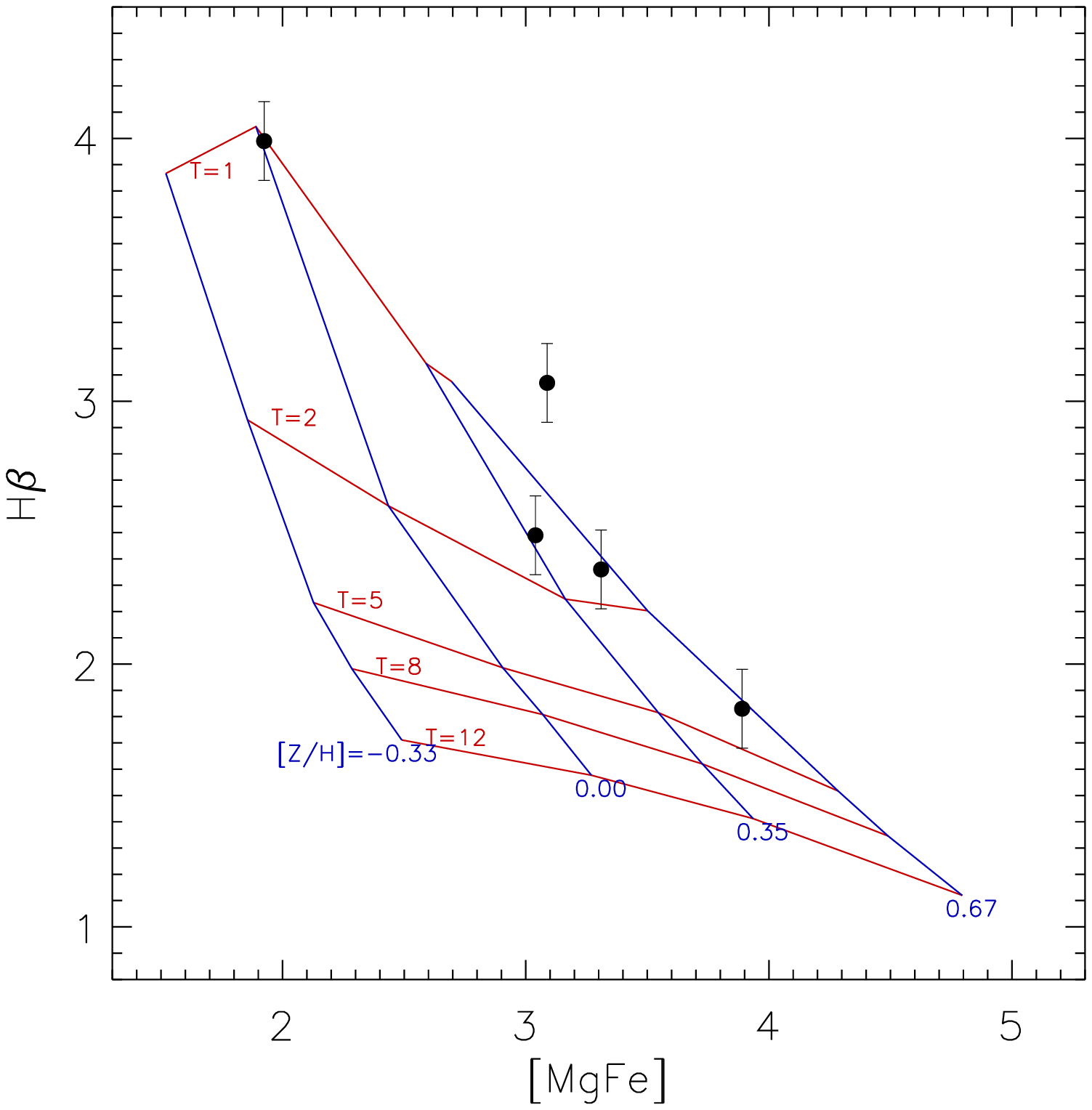} \\
 (a)&(b)\\
\end{tabular}
\caption{
The diagnostic diagrams `index--index' for the nuclei of five off-central galaxies
of the NGC 524 group (black points with the error bars). Small signs connected
by dashed lines on the left plot and solid red lines on the right plot represent
the model age sequences from \cite{thomod}; the age labels are in Gyrs; model repers
(small signs on the left plot and blue line bases on the right plot) are marked by
the total metallicity [Z/H].}\label{nuclei}
\end{center}
\end{figure*}

Table~3 gives the Lick indices and the stellar population parameters, derived 
from them, for the nuclei of the studied galaxies. Figure~6 demonstrates a comparison
of the observed indices with the models from \cite{thomod}. The first thing that 
catches the eye is that in four of five galaxies, the stellar nuclei are very young, 
with an average age of the stellar population of 1–2 Gyr. Secondly, in four of five 
galaxies (though not the same) the metallicity of the stellar population in the
nuclei is very high, with the total metal abundance 3-5 times higher than the solar. 
Obviously, we see the consequences of the recent highly efficient starforming bursts
in the nucleus, and the chemically decoupled nuclei of galaxies they generated. The
magnesium-to-iron ratio in three out of four young nuclei is solar, and only in NGC~518, 
where we noted a larger compactness of the magnesium index peak
compared with that of the iron (Fig. 4), this ratio is one and a half times greater
than solar. That means that in these three nuclei the burst of star formation has been 
long enough, lasting for at least 2-3 Gyr. It was perhaps more brief in the nucleus
of NGC~518, though it was going on for the same 2--3 Gyr in the immediate vicinity 
of the nucleus, already starting from the radius of 2'' (300 pc).

\begin{table*}
\caption[ ] {Lick indices and stellar population parameters in the bulges}
\begin{center}
\begin{tabular}{l|c|c|c|c|c|c|}
\hline\noalign{\smallskip}
Galaxy & H$\beta$ &  Mgb & $\langle \mbox{Fe} \rangle$ 
& T, Gyr & [Z/H] & [Mg/Fe] \\
\hline\noalign{\smallskip}
NGC 502 & 1.95 & 4.13 & 2.82 & 4 & $+0.3$ & $+0.19$  \\
NGC 509 & 1.90 & 3.06 & 3.02 & 5 & $+0.1$ & $-0.13$ \\
NGC 516 & 1.96 & 2.65 & 2.84 & 6 & $-0.1$ & $-0.19$ \\
NGC 518 & 2.23 & 3.12 & 2.78 & 3 & $+0.1$ & 0.0 \\
NGC 532 & 2.06 & 3.42 & 2.68 & 4 & $+0.1$ & $+0.10$ \\
\hline
\end{tabular}
\end{center}
\end{table*}

Table 4 lists the Lick indices and the stellar population parameters derived 
from them for the bulges of the galaxies under consideration; the bulges taken 
in the rings between $R=4''$ and $R=6''$ (0.65--1.0 kpc radius). Figure 7 compares
these data to the models from \cite{thomod}. The scatter of the magnesium-to-iron 
abundance ratio in the stellar populations of the bulges is significant: it may
be either higher than solar (in the bulge of NGC 502), or significantly lower than 
the solar, as in the bulges of NGC 509 and NGC 516. If [Mg/Fe]$>0$ is known to be 
typical for rather large spheroids \cite{wor92,trager00b}, [Mg/Fe]$<0$ is on the 
other hand a very rare case among the galaxies, and is found mainly in irregular
dwarfs of the Local Group. The magnesium underabundance can be treated in the frame
of the chemical evolution theory as a signature of intermittent character of star 
formation, with a small number of bursts separated by long, 2--3~Gyr lasting, quiescent 
periods \cite{gilmorewyse}. Meanwhile, the mean ages of stellar populations 
in all the five bulges of the NGC524 group galaxies are confined to a fairly narrow 
range of values, those between 3 and 5 Gyr, if for NGC 509 and NGC 516 we use 
the models from \cite{thomod} with deficiency of magnesium relative to iron (Fig. 7c).

\begin{figure*}[h]
\begin{tabular}{c c c}
 \includegraphics[width=5cm]{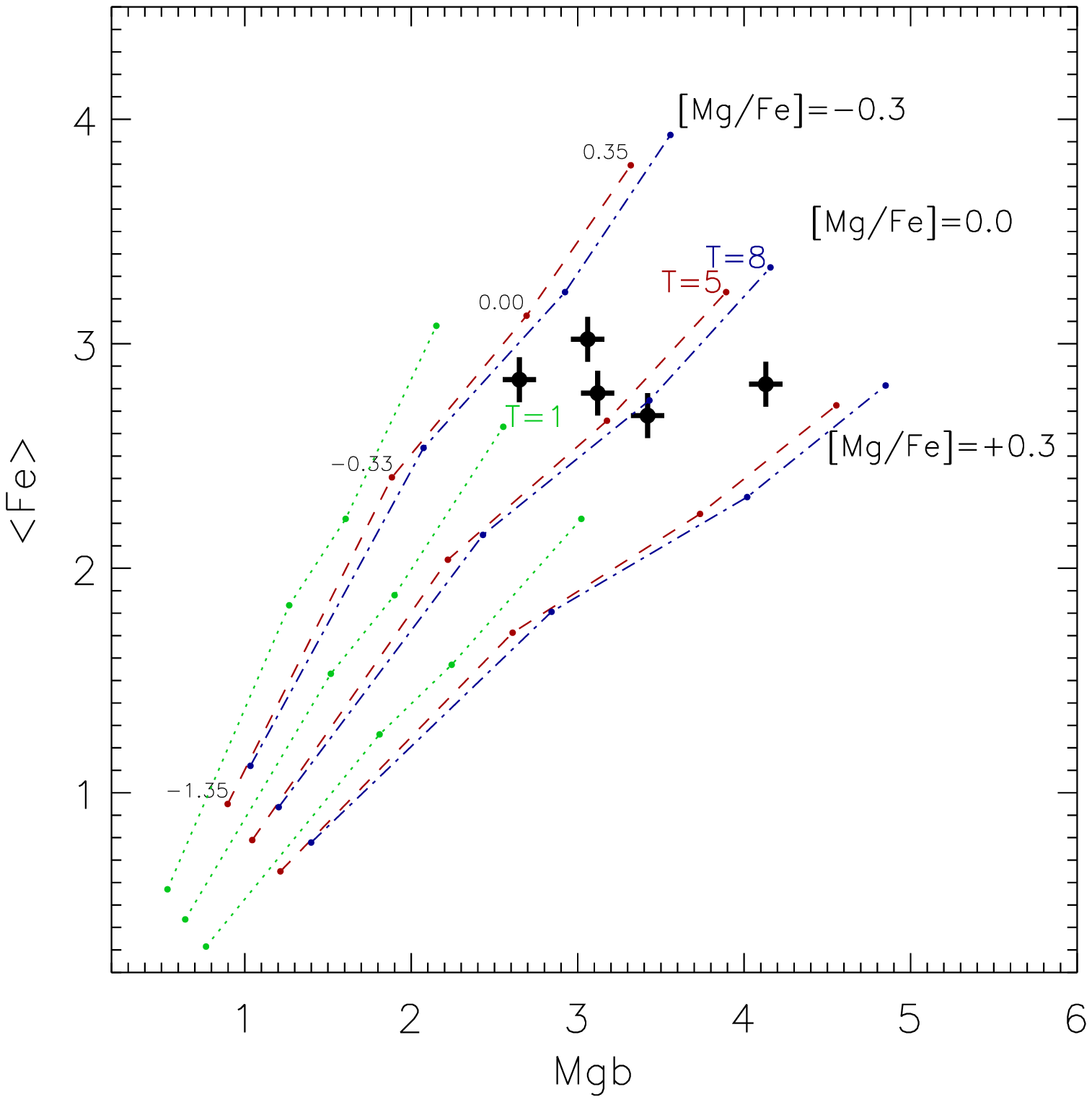} &
 \includegraphics[width=5cm]{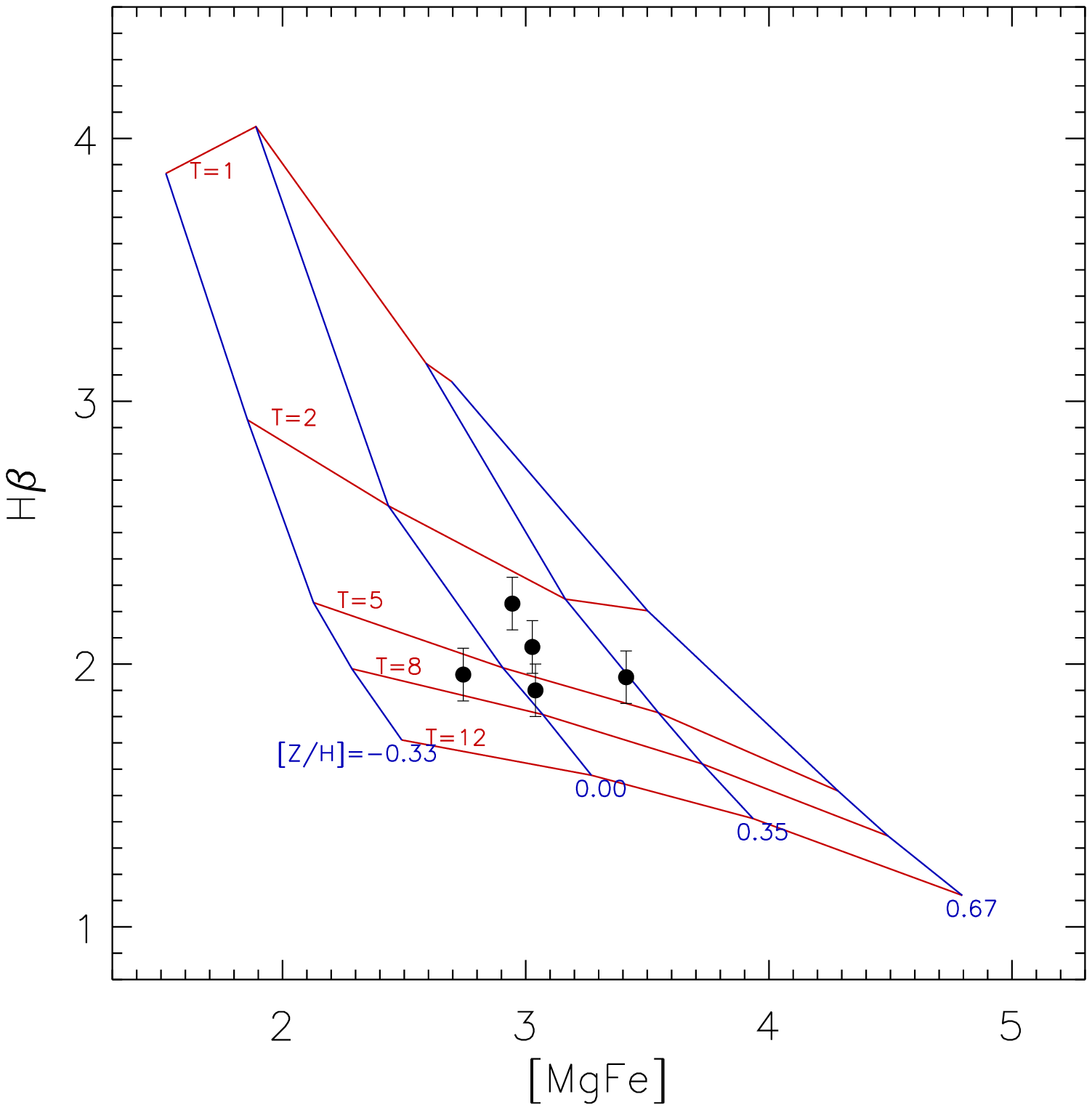} &
 \includegraphics[width=5cm]{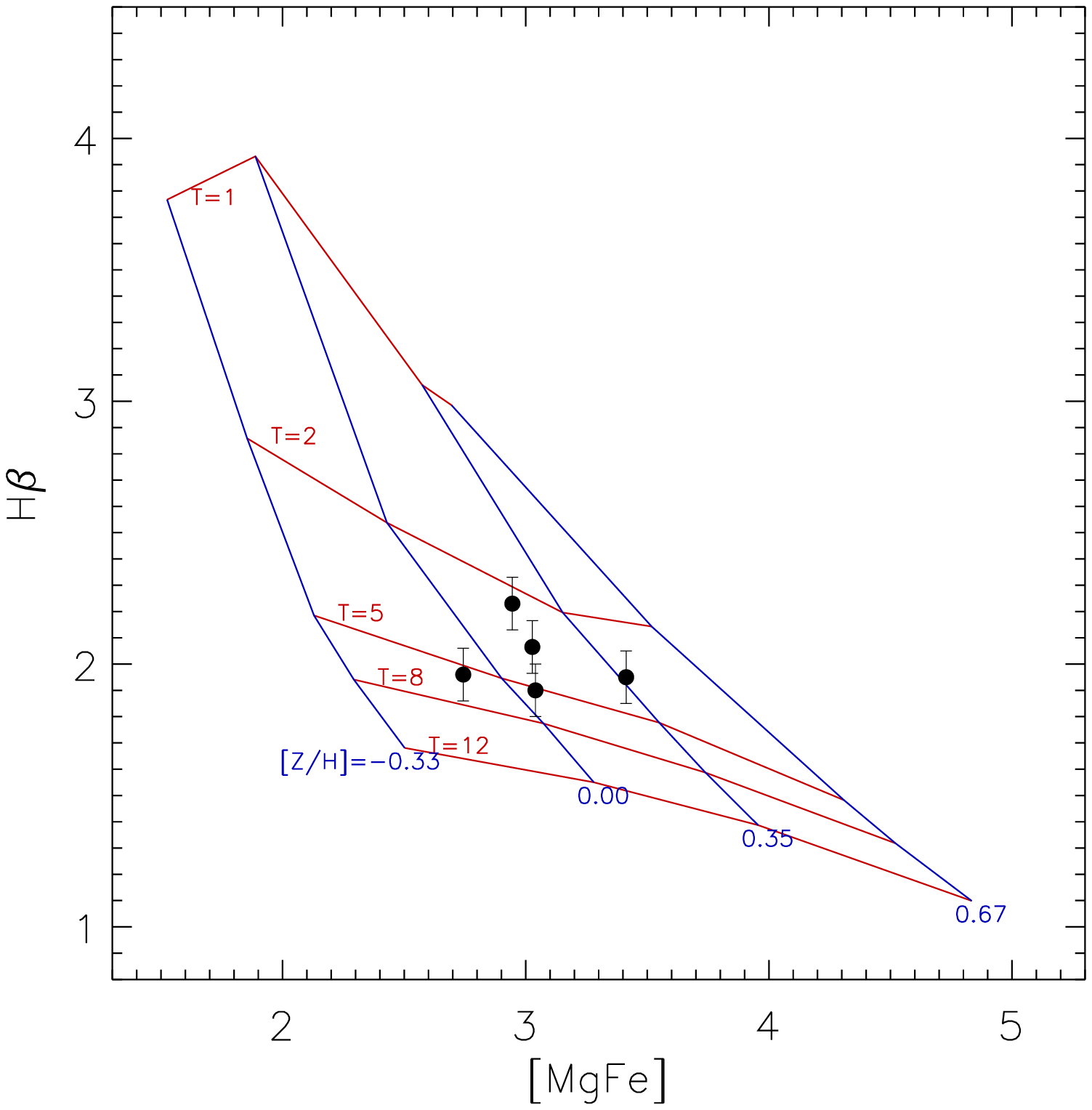} \\
 (a) & (b) & (c) \\
\end{tabular}
\caption{
The diagnostic diagrams `index--index' for the bulges taken in the rings between 
the inner and outer radii of $4^{\prime \prime}$  and $6^{\prime \prime}$ 
(black points with the error bars). Small signs connected
by dashed lines on the left plot and solid red lines on the right plots represent
the model age sequences from \cite{thomod}; the age labels are in Gyrs; model repers
(small signs on the left plot and blue line bases on the right plots) are marked by
the total metallicity [Z/H]. Two age-diagnostic diagrams ``H$\beta$--[MgFe]'' 
are plotted for two magnesium-to-iron ratios: (b) [Mg/Fe]$=0.0$, 
(c) [Mg/Fe]$=-0.3$.}\label{bulges}
\end{figure*}

The accuracy of the ages of stellar populations calculated for the bulges 
of five studied galaxies is about 1 Gyr if the galaxy age is of 3--4 Gyr, 
and about 2 Gyr for the ages of 5--6 Gyr. That means that the bulges of all 
the five studied galaxies have approximately the same mean age of the stellar 
populations within the precision claimed.

\subsection{Relic circumnuclear rings}

In addition to the bulges and unresolved stellar nuclei --
two separate stellar subsystems that were found to have
substantially different star formation histories in most
of the galaxies studied here, -- we can also point out such curious features
as relic circumnuclear rings of past star formation. As we already noted,
in NGC 518, at the radius of 300 pc, the iron index suddenly jumps up, 
while the magnesium index monotonously decreases, so that the Mg/Fe ratio 
in the nucleus and its nearest vicinity differs by a factor of one and a half. 
This can be explained by a prolonged star formation in the ring compared 
with a brief starburst in the geometric center of the galaxy. However, 
in NGC 518 this difference in the duration of star formation along the
radius is visible only through the chemical composition of stars and is not
revealed in the estimates of the mean ages of stellar populations. 
The mean age of stars in NGC 518 marginally increases from 2 to 3 Gyr along the radius
so being constant within the estimate accuracy. In NGC 532, the circumnuclear ring 
of recent star formation can be seen by a `naked eye'. Figure 8a shows 
the H$\beta$--[MgFe] diagram for NGC 532: the azimuthally averaged measurements 
of the Lick indices along the radius of the galaxy, starting from
the nucleus, then in the ring at $R=2''$ and further on with a step of one
arcsecond. It can be seen how with the distance from the nucleus the age 
of the stellar population drops initially by a few hundred million years, and then
abruptly increases, while in the bulge at the distances of 0.6–1.0 kpc from the center  
it practically does not vary. We can see that the youngest stellar population
in the center of NGC 532 is concentrated in the ring at the radius of 2''. 
Curiously, this radius is distinguished both photometrically and kinematically.
Figure 8b shows the variation of the isophote major axis orientation along 
the radius (calculated and published by us \cite{n524phot}) in comparison to
the kinematical major axes, determined from the velocity fields of stars and 
ionized gas; the bottom plot shows the variationa of the isophote ellipticity.
The $R = 2''- 3''$ point is photometrically distinct: at this radius, 
the ellipticity of the isophotes is maximal, and the major axes of the stellar 
component, both photometric and kinematical ones, simultaneously turn
to larger position angles. Since the kinematical major axis of the stellar 
component is turned with respect to the line of nodes in the same direction 
as the photometric major axis, we can interpret the compact elongated stellar 
structure  in the center of NGC 532 as an inclined circumnuclear stellar disk, 
which has been formed, as we have seen from the Lick indices analysis, rather recently,
within 1~Gyr from our epoch.

\begin{figure*}[h]
\begin{center}
\begin{tabular}{c c}
 \includegraphics[width=6cm]{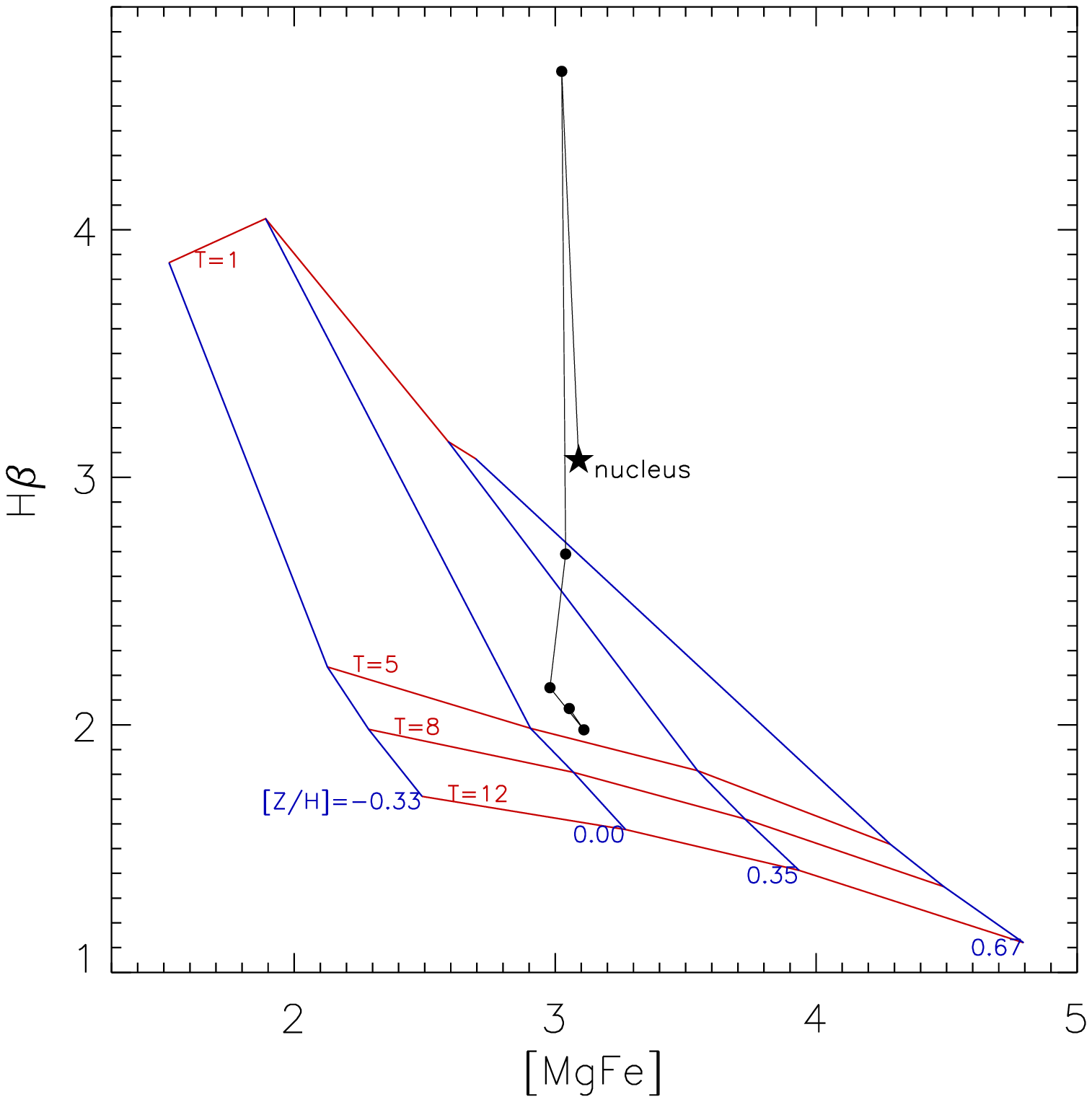} &
 \includegraphics[width=6cm]{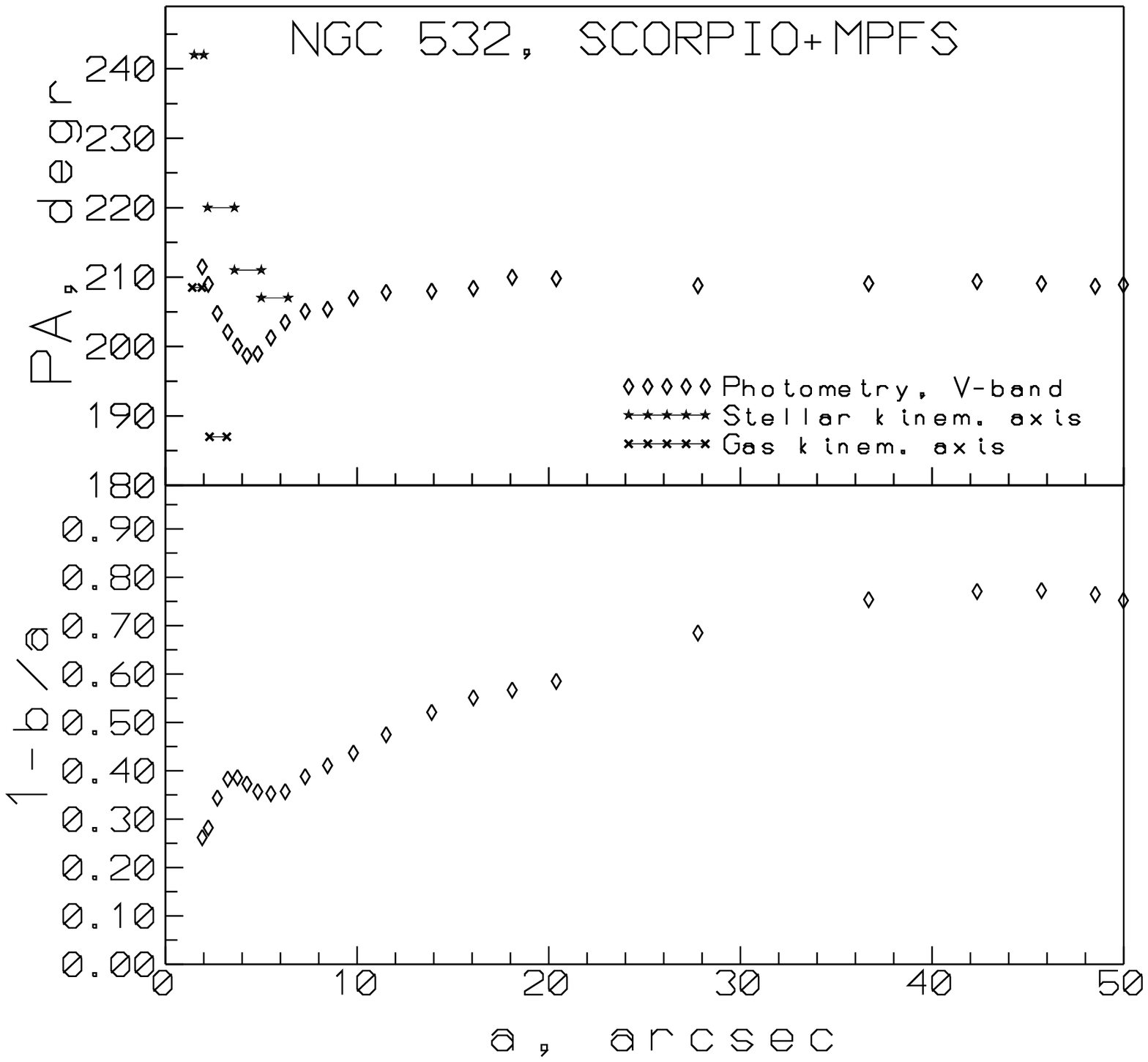} \\
 (a)&(b)\\
\end{tabular}
\caption{
Circumnuclear ring in NGC~532: {\it (a)} Diagnostic diagram ``H$\beta$--[MgFe]'' 
for the nucleus (star) and off-nuclear Lick index measurements (black dots) starting from 
$R=2^{\prime \prime}$, and with a radial step of $1^{\prime \prime}$ further on; the model
sequences are the same as in Figs.~ 6b and 7b;
{\it(b)} The results of the isophote analysis of the $V$-band SCORPIO image
of NGC~532 \cite{n524phot}; the upper plot compares the orientations of the
photometric and kinematical major axes; the latters are determined for the
stellar and gas velocity fields by exploring the MPFS data.}\label{n532ring}
\end{center}
\end{figure*}

\section{CONCLUSIONS AND DISCUSSION}

We have investigated the properties of stellar populations
in the central regions of five early-type disk galaxies of the NGC524 group 
via the method of panoramic spectroscopy, by using the Multi-Pupil Fiber Spectrograph (MPFS),
mounted at the prime focus of the Russian 6-m telescope. The central galaxy of the group, 
a giant S0 NGC524, was investigated by us by using the same instrument set up 
earlier \cite{n524}. We found there a homogeneously old ($T > 10$~Gyr)
stellar population in all its substructures -- in the stellar nucleus, in 
the bulge, and in the multi-tiered large-scale disk [15, 33]. In contrast 
with the central galaxy, the other galaxies of the group, as it turned out,
have intermediate-age populations in the bulges and young stellar populations 
in the chemically decoupled stellar nuclei. Interestingly, the mean (SSP-equivalent) 
age of the stars in the bulges is about the same in all the five galaxies, 
$\sim 3-5$~Gyr, and the mean stellar age of the nuclei is approximately
the same in four of the five galaxies studied, $T_{SSP} \sim 1-2$~Gyr. 
Only the S0-galaxy NGC~502, the most distant from the group center, did not undergo
a separate burst of star formation in the nucleus: its nucleus and bulge have 
the same age of 4~Gyr. Such a synchronous evolution of the galaxies in the
group implies the decisive role of environment in stimulating the inner events 
in the lives of galaxies, and the bursts of star formation in their centers,
first of all. In the NGC~524 group, no hot-gas medium is present over the entire 
volume of the group; only a compact X-ray halo was registered around the NGC~524 
itself. It is therefore evident that the gravitational interactions play the defining
role here. The time scales of building up the central regions of galaxies that we
have identified for the NGC~524 group members, agree with the concordant cosmological 
simulations of the hierarchical clustering of matter in the Universe. Indeed,
in LCDM cosmological models, massive clusters/groups of galaxies, $M > 10^{13}M_{odot}$, 
become gravitationally bound near the redshift one. For example, in a
recent study by \cite{kaufmann} the first ``satellites'' (non-central
galaxies of a group) cross the virial group radius at the redshift of 1.5 (
9 Gyr ago), and the most recent -- around $z = 0.7$ (6 Gyr ago).
Falling within the virial radius, the galaxies experience the gravitational 
influence of both the potential of the entire group \cite{villa}, and the 
tidal effects from the other group members during close passages \cite{bekkicouch}.
These gravitational interactions distort the axial symmetry of the galactic disks, 
accreted by the group, leading to the formation of bars, and the bars, in their turn,
redistribute the gas in the disks, forcing it to accumulate in the centers of galaxies. 
Further on, the bars may dissolve under the gravitational effect of the
central concentration of mass. As a result, secondary bursts of star formation are 
expected exactly in the central kiloparsec, which should lead to the growth
of bulges (and to a decrease of the mean age of the bulge stars), while in the 
galaxies that have not had prominent bulges before -- just to the formation of 
central spheroidal components, or pseudobulges. Indeed, at $z = 0.4 - 0.7$ in a 
dense environment of clusters and groups the blue disk galaxies are 
dominating \cite{wilman,just}, which may be mistreated as spiral galaxies
that are only approaching their transformation into S0s. In effect, these 
may be the disc galaxies including the S0s, that are in the process of a vigorous
build-up of the bulges immediately after the accretion into a dense 
environment \cite{burst_s0}. This process is observed to stop at $z < 0.4$, 
which is 3--4 Gyr ago, what is consistent with the bulge ages found by us in
the disk galaxies of the NGC~524 group. Further, about 1-2~Gyr ago, the galactic 
nuclei experienced the secondary bursts of star formation, more compact, and obviously 
with more modest consequences than those that resulted in building-up the bulges. 
Interestingly, they are also synchronized. They are likely to be related with the
dynamic evolution of the group as a whole: the crossing time of the group
is approximately 2-3~Gyr. Around the large galaxies of the NGC~524 group we 
have found a significant number of blue dwarf galaxies, obviously rich in
gas \cite{n524phot}. Any passage of a large galaxy through the center
of the group may possibly provoke multiple accretion events by disturbing
the orbits of dwarf companions, and this would supply the fuel for a nuclear 
burst of star formation, which had to be quickly and efficiently burnt out
(the lenticular galaxies NGC~516 and NGC~509, where we found young stellar nuclei, 
are completely gas--free, according to \cite{sengupta}).

\section{ACKNOWLEDGMENTS}

The telescope operation and spectrograph managing was supported 
by the Ministry of Education and Science of Russian Federation (state
contracts no. 16.552.11.7028 and 16.518.11.7073).
We thank A. V. Moiseev and A. A. Smirnova for the
observational support with the MPFS. During the analysis
of the results, we have used the Lyon-Meudon Extragalactic Database (LEDA),
maintained by the LEDA team of the CRAL at the Lyon Observatory (France), 
and the NASA/IPAC (NED) database of extragalactic data, managed by
the Jet Propulsion Laboratory at Caltech under the state contract with 
the National Aeronautics and Space Administration (USA). This study of 
lenticular galaxies was supported by the RFBR grant no. 10-02-00062a.

\end{document}